\documentclass{aa}
\usepackage{graphicx}
\newcommand{\ori}{$\alpha$~Orionis\,\,\,}
\newcommand{\her}{$\alpha$~Herculis\,\,\,}
\begin{document}
\title{Interferometric observations of the supergiant stars 
$\alpha$~Orionis and $\alpha$~Herculis with FLUOR at IOTA\thanks{based 
on observations collected at the IOTA interferometer, Whipple 
Observatory, Mount Hopkins, Arizona.}}

\author{G. Perrin \inst{1} \and S.T. Ridgway \inst{2}\and V. Coud\'e 
du Foresto \inst{1} \and B. Mennesson \inst{3} \and W.A. Traub 
\inst{4} \and M.G. Lacasse \inst{4}}

\offprints{G. Perrin}

\institute{Observatoire de Paris, LESIA, UMR 8109, F-92190 Meudon, 
France \and National Optical Astronomy Observatories, Tucson, AZ 
85726-6732, USA \and Jet Propulsion Laboratory, California Insititute 
of Technology, MS 306-388, 4800 Oak Grove Drive, Pasadena, CA 91109, 
USA \and Harvard-Smithsonian Center for Astrophysics, Cambridge, MA 
02138, USA}

\titlerunning{Interferometric observations of
$\alpha$~Orionis and $\alpha$~Herculis with FLUOR at IOTA}

   \date{Received / accepted }

    \abstract{ We report the observations in the K band of the red supergiant star $\alpha$~Orionis and of the bright giant star $\alpha$~Herculis with the FLUOR beamcombiner at the IOTA interferometer.  The high quality of the data allows us to estimate limb-darkening and derive precise diameters in the K band which combined with bolometric fluxes yield effective temperatures. In the case of Betelgeuse, data collected at high spatial frequency although sparse are compatible with circular symmetry and there is no clear evidence for departure from circular symmetry.  We have combined the K band data with interferometric measurements in the L band and at 11.15\,$\mu$m.  The full set of data can be explained if a 2055\,K layer with optical depths $\tau_{K}=0.060\pm0.003$, $\tau_{L}=0.026\pm0.002$ and $\tau_{11.15\mu m}=2.33\pm0.23$ is added 0.33\,$R_{\star}$ above the photosphere providing a first consistent view of the star in this range of wavelengths.  This layer provides a consistent explanation for at least three otherwise puzzling observations: the wavelength variation of apparent diameter, the dramatic difference in limb darkening between the two supergiant stars, and the previously noted reduced effective temperature of supergiants with respect to giants of the same spectral type.  Each of these may be simply understood as an artifact due to not accounting for the presence of the upper layer in the data analysis.  This consistent picture can be considered strong support for the presence of a sphere of warm water vapor, proposed by \cite{tsuji2000} when interpreting the spectra of strong molecular lines.
 }

   \maketitle
%

\section{Introduction}
Because of their large luminosity supergiant
stars have extended atmospheres, with resultant low surface gravity and
large pressure scale height.  The cool supergiants are usually slightly
variable in brightness, color, polarization, and radial velocity,
suggestive of significant dynamic and radiative activity and
inhomogeneities.  However, the variablity is small enough that static model
atmospheres are normally considered adequate for abundance studies.  Owing
to the very extended atmospheres, and in some cases circumstellar shells,
it is difficult to determine even some very basic parameters for these stars. 
Here, we address the question of the effective temperatures. The near
infrared wavelength regime offers several advantages for the study of cool
supergiants - it is less obscured by scattering and line blanketing
(problems in the visible) and by dust (detected in the mid-infrared). \\

\begin{figure*}
    \begin{center}
	\hbox{    
	      \includegraphics[width=9cm]{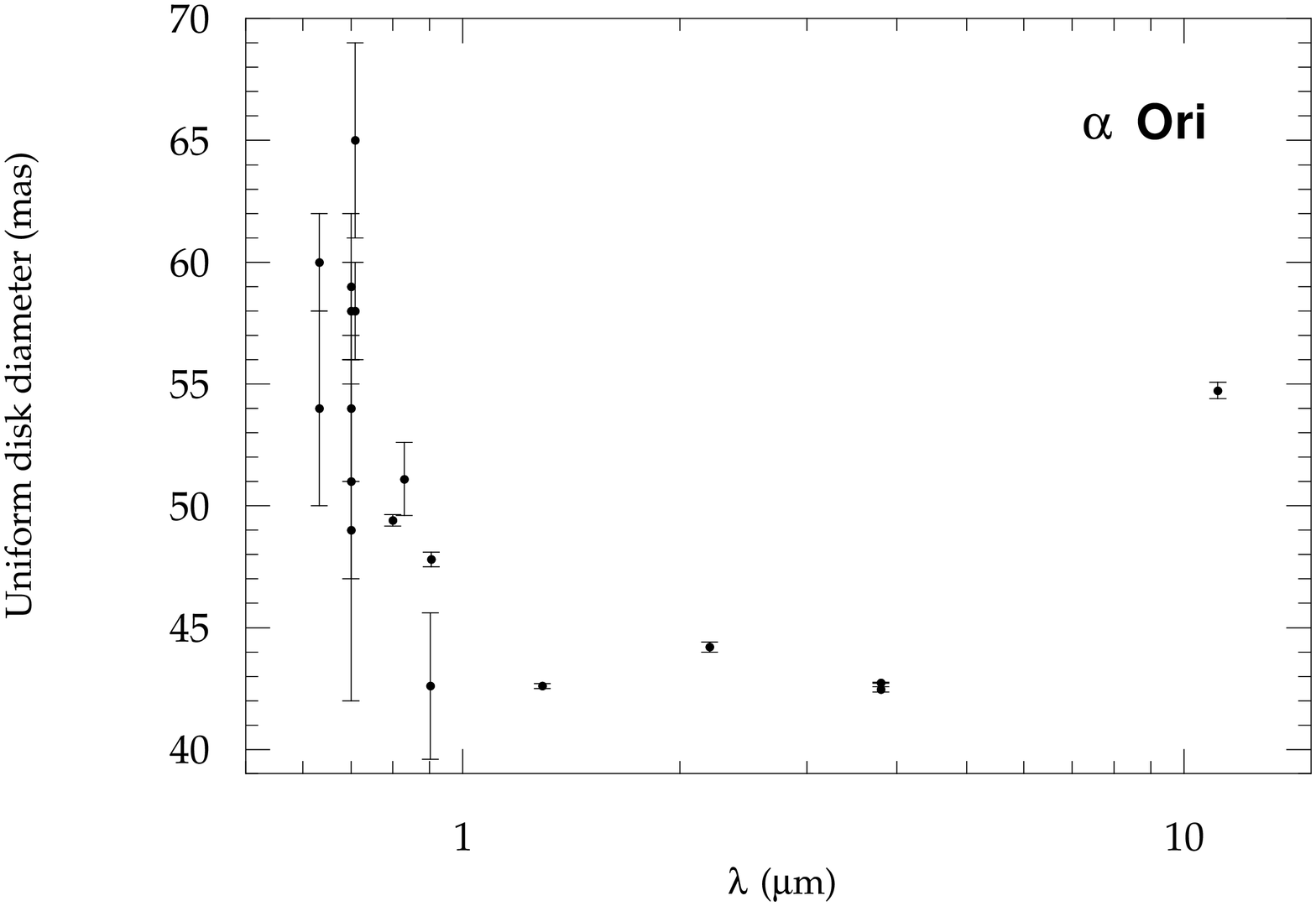}
              \includegraphics[width=9cm]{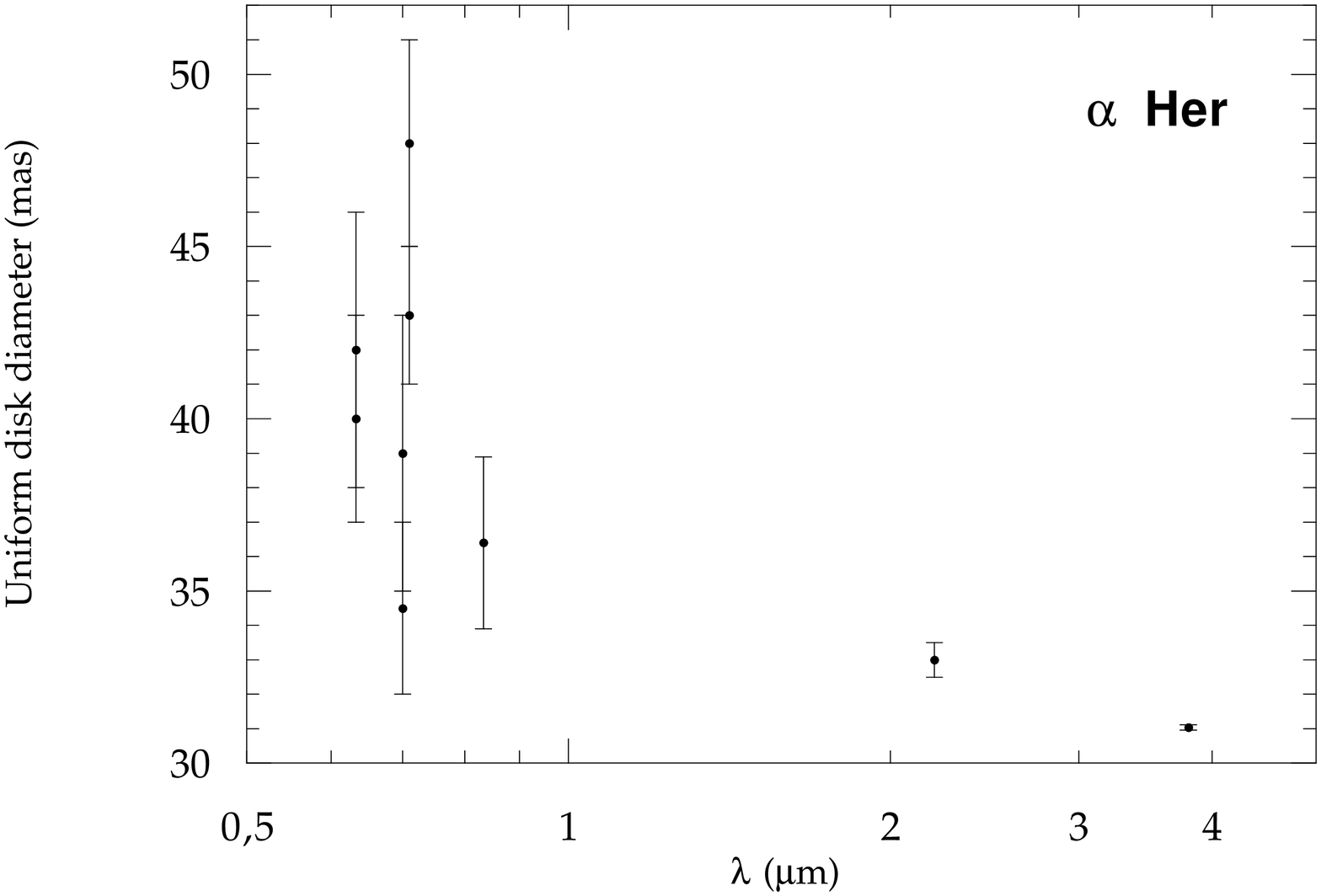}
      } \caption{Uniform disk diameters measured for $\alpha$~Orionis 
      and $\alpha$~Herculis.  Values are reported in \cite{benson91, 
      mozurkewich91, dyck92, tuthill97, burns97, mennesson99, 
      young2000, weiner2000, chagnon02}.}
         \label{fig:UDvslambda}
   \end{center}
     \end{figure*}

$\alpha$~Orionis is a bright, prototypical example of the cool 
supergiant class.  It is classified M1-2Ia-Iab, which places it firmly 
in the supergiant category.  An infrared excess and extended dust 
shell confirm substantial mass loss.  $\alpha$~Orionis has been 
observed several times with high angular resolution techniques and in 
several optical bandpasses.  The derived angular diameters are plotted 
against wavelength in Fig.~\ref{fig:UDvslambda}.  The observed angular 
diameters are significantly larger in the visible, decrease to a 
minimum in the near-infrared, and then increase again in the 
mid-infrared.  Some surface structure, similar to large spots, has been 
seen at several wavelengths in the red (\cite{tuthill97}, 
\cite{young2000}).  These spots have not been seen at other nearby 
wavelengths.

We have observed $\alpha$~Orionis in the K band with FLUOR at IOTA in 1996
and 1997.  We will present these observations in the next section and
explain the data reduction procedure.  In the following sections we will
derive uniform disk, limb darkened and linear diameters.  We will compare
our results to the previous observations and derive a simple model for the
wavelength dependence of diameters in section~\ref{sec:discussion}. We also
present similar observations of $\alpha$~Herculis.  $\alpha$~Herculis has
been classified M5Ib-II, indicative of a somewhat lower luminosity. It does
not have an infrared excess. Previously determined diameters are also 
plotted in  Fig.~\ref{fig:UDvslambda}.  \\

\section{Observations and data reduction}
\begin{table*}
    \begin{center}
      \caption[]{Log of observations}
         \label{tab:obs}
         \begin{tabular}{lccccccc}
            \hline
            \noalign{\smallskip}
            Source      &  Date & Band & Spatial frequency & Position angle & 
            Visibility & Reference 1 & Reference 2\\
	                &       & & (cycles/arcsec)   & $(^o)$ & & &\\
            \noalign{\smallskip}
            \hline
	    $\alpha$~Orionis & 18/02/1996 & K & 44.08 &  -    & $0.0843\pm0.0024$ & HD18884  & HD94705 \\
	                     & 9/11/1996  & K & 24.43 & 63.01 & $0.1253\pm0.0016$ & HD18884  & HD42995 \\
                             & 9/11/1996  & K & 23.34 & 71.96 & $0.1847\pm0.0024$ & HD18884  & HD42995 \\
			     & 9/11/1996  & K & 22.70 & 80.47 & $0.2045\pm0.0020$ & HD44478  & HD44478 \\
			     & 9/11/1996  & K & 22.50 & 85.91 & $0.2232\pm0.0018$ & HD44478  & HD44478 \\
			     & 22/02/1997 & K & 64.96 & 79.06 & $0.0510\pm0.0010$ & HD61421  & HD61421 \\
			     & 9/03/1997  & K & 42.81 & 69.89 & $0.0949\pm0.0047$ &    -     & HD61421 \\
             \noalign{\smallskip}
            \noalign{\smallskip}
	    $\alpha$~Herculis& 17/04/1996 & K & 35.67 & 53.97 & $0.0800\pm0.0038$ & HD89758  & HD186791\\
			     & 17/04/1996 & K & 35.26 & 59.28 & $0.0894\pm0.0022$ & HD89758  & HD186791 \\
			     & 21/04/1996 & K & 46.57 & 106.31& $0.0954\pm0.0068$ & HD130144 & - \\
			     & 6/03/1997  & K & 35.05 & 60.90 & $0.1072\pm0.0028$ & HD130144 & -  \\
                             & 6/03/1997  & K & 35.51 & 57.07 & $0.0927\pm0.0027$ & HD130144 & - \\
            \noalign{\smallskip}
            \noalign{\smallskip}
            \hline
         \end{tabular}
	 \end{center}
   \end{table*}
   
\begin{figure}
    \begin{center}
\includegraphics[width=9cm]{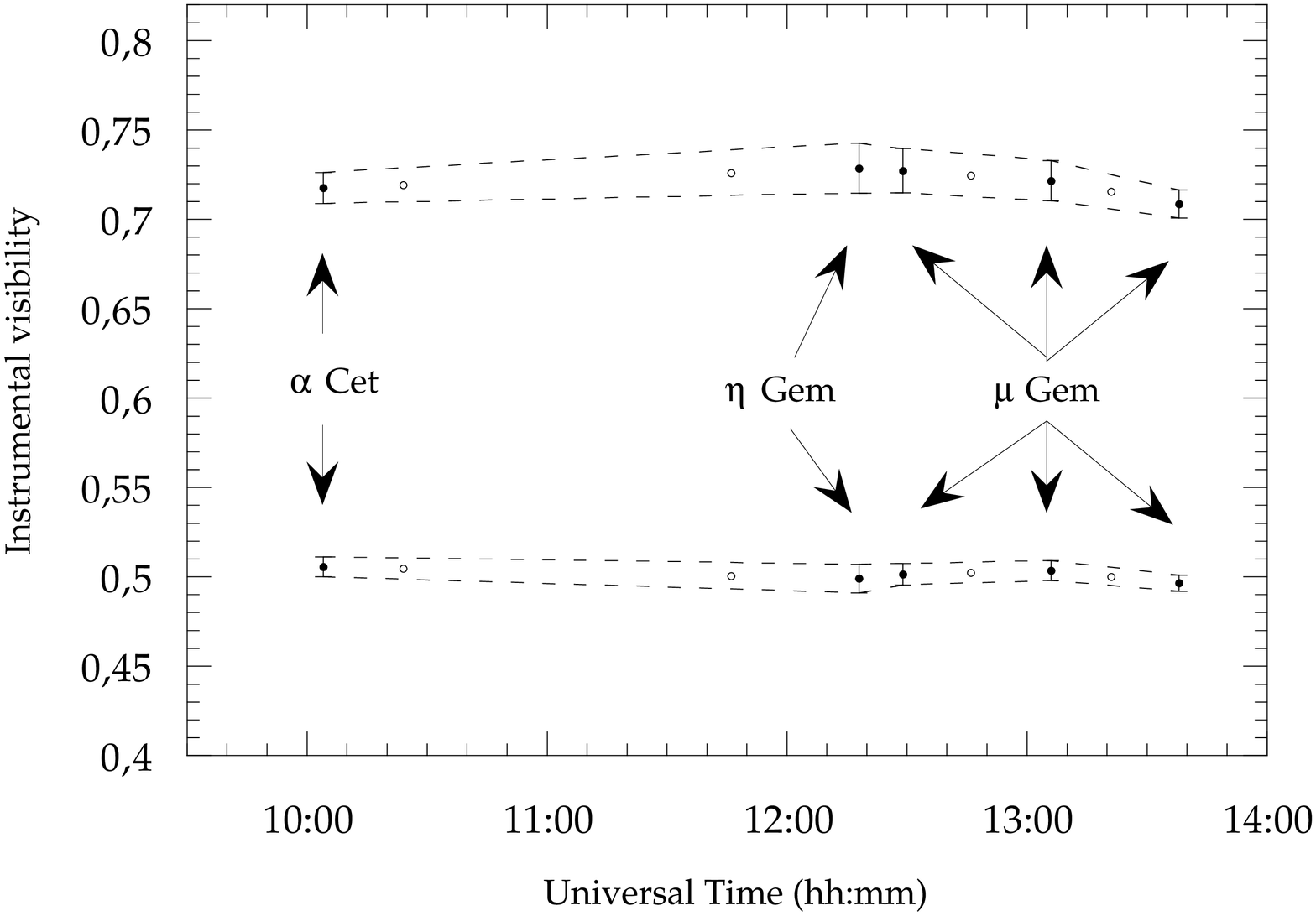}
      \caption{Instrumental visibility on November 9, 1996 for the two 
      interferometric channels (full circles).  Open circles are the 
      interpolated instrumental visibilities at the time Betelgeuse 
      was observed.}
         \label{fig:transfert}
   \end{center}
     \end{figure}

\begin{table}
      \caption[]{Reference sources}
         \label{tab:cal}
         \begin{tabular}{lccc}
            \hline
            \noalign{\smallskip}
            HD number & Source  &  Spectral type & Uniform disk  \\
	                 & name &                & diameter (mas)\\
            \noalign{\smallskip}
            \hline
	    HD18884  & $\alpha$~Cet & M1.5 III & $12.66\pm0.36$$^{\mathrm{a}}$\\
	    HD94705  & 56 Leo       & M5.5 III & $8.66\pm0.43$$^{\mathrm{b}}$\\
	    HD42995  & $\eta$~Gem   & M3 III   & $10.64\pm0.53$$^{\mathrm{b}}$\\
	    HD44478  & $\mu$~Gem    & M3 III   & $13.50\pm0.15$$^{\mathrm{c}}$ \\
	    HD61421  & $\alpha$~CMi & F5 IV    & $5.25\pm0.26$$^{\mathrm{b}}$\\
	    HD89758  & $\mu$~UMa    & M0 III   & $8.28\pm0.41$$^{\mathrm{b}}$\\
	    HD186791 & $\gamma$~Aql & K3 II    & $6.99\pm0.35$$^{\mathrm{b}}$\\
	    HD130144 & HR 5512      & M5 III   & $8.28\pm0.41$$^{\mathrm{b}}$\\
            \noalign{\smallskip}
            \noalign{\smallskip}
            \hline
         \end{tabular}
\begin{list}{}{}
\item[$^{\mathrm{a}}\,\,$\cite{cohen96}]
\item[$^{\mathrm{b}}$ Photometric estimate]
\item[$^{\mathrm{c}}\,\,$\cite{benedetto87}]

\end{list}
   \end{table}

The stars have been observed in 1996 and 1997 with the IOTA 
(Infrared-Optical Telescope Array) interferometer located at the 
Smithsonian Institution's Whipple Observatory on Mount Hopkins, 
Arizona (\cite{traub98}).  Several baselines of IOTA have been used to 
sample visibilities at different spatial frequencies.  The data have 
been acquired with FLUOR (Fiber Linked Unit for Optical Recombination) 
in the K band.  Beam combination with FLUOR is achieved by a 
single-mode fluoride glass triple coupler in the K band.  The fibers 
spatially filter the wavefronts corrugated by the atmospheric 
turbulence.  The phase fluctuations are traded against photometric 
fluctuations which are monitored for each beam to correct for them a 
posteriori.  At the time of the reported observations the modulation 
of the optical path difference was produced by scanning through the 
fringe packet with the IOTA short delay line and the four signals were 
detected with InSb single-pixel detectors (\cite{perrin96}).  The 
limiting magnitude was then K=0.  The accuracy on visibility estimates 
measured by FLUOR is usually better than 1\% for most sources 
(\cite{perrin98}).\\
The log of the observations is given in Table~\ref{tab:obs}.  The 
second column is the UT date of the observations.  The next two 
columns are the spatial frequency and the position angle of the 
spatial frequency vector.  The next column is the visibility 
and the 1\,$\sigma$ error.  The last two columns are the HD numbers of 
the reference sources observed just before and just after the science 
target to estimate the instrumental visibility.  The characteristics 
of the reference sources (spectral type and uniform disk diameter) are 
listed in Table~\ref{tab:cal}.  Diameters have either been measured or 
derived from photometric and spectroscopic scales (\cite{cohen96}, 
\cite{perrin98}).  Each line in Table~\ref{tab:obs} corresponds to two 
batches of data: the signals acquired on-source (the fringes) 
representing a collection of at least 100 scans; the signals acquired 
off-source to estimate the contribution of the detector to the noise 
(same number of scans) in the K band.  These two batches are called an 
observation block.  The same sequences are repeated for the reference 
stars.\\

The procedure explained in \cite{foresto97} has been applied to all 
sources independently to estimate the average constrast of the fringe 
packet for each observation block.  The expected visibility of the 
calibrators is then computed at the time they were observed.  
The instrumental visibility 
is then interpolated at the time when the science targets have been 
observed.  Division of the fringe contrast of the science targets by 
the interpolated instrumental visibility provides the final visibility 
estimate.  This procedure has been explained in \cite{perrin98}.  It 
has been recently refined (\cite{perrin02a}) to take into account 
correlations in fringe contrast estimates and transfer function 
estimates in the computation of error bars.  In the special case of 
the February 18$^{th}$, 1996 K band observation, the triple coupler was not 
available and we had to use a single coupler instead.  The only
difference is that no photometric signals are available.  Instead, the
low frequency part of the interferograms is used to estimate the
photometric fluctuations.  The calibration is therefore less accurate
than with the triple coupler and the possibility of bias in the final
visibility estimate must be considered.  However, the bias being
proportional to the visibility is more important when the  average
intensities in the two interferometric arms are not well balanced.  In
the case of this particular observation, the visibility is low and the
two arms were well balanced. For these reasons the bias is negligible and
the data are retained. \\


A manual correction of the bias of the squared fringe contrast by the 
source photon noise has been added to the regular procedure of data 
reduction for the K band data.  At the time when InSb detectors were 
used, the sensitivity of FLUOR was so low (K=0) that the bias due to 
photon noise could usually be  neglected.  
Yet, on very bright sources like \ori and $\alpha$~Herculis, the bias 
is larger than the tiny error bars achieved and its correction is 
mandatory for best results.  The general expression of the photon noise 
bias is well 
described in \cite{goodman85}.  This expression can be adapted in the 
case of FLUOR giving the theoretical value of the bias that has to be 
removed from the data (\cite{perrin02b}).  When the conversion factor 
between ADUs (or detector output voltage) and photons is well known, the 
removal of the bias is easy 
and can be automated.  This is not unfortunately the case for the data 
presented in this paper and this is why the procedure had to be 
manual. \\ \\

To illustrate the great stability of the FLUOR measurements in the K 
band, the instrumental visibility for November 9, 1996 has been 
plotted on Figure~\ref{fig:transfert} for the two interferometric 
channels.  The dashed lines represent the $1\,\sigma$ lower and upper 
limits.  Three different calibrators have been used and the transfer 
function is stable to better than $1\,\sigma$ on a time scale of 2 
hours.  The interpolation of the transfer function to the time of the 
observation of the science target (open circles) therefore yields a 
very accurate estimate of the visibility transfer function.

\section{Uniform disk diameters}
\begin{figure*}
    \begin{center}
\includegraphics[width=18cm]{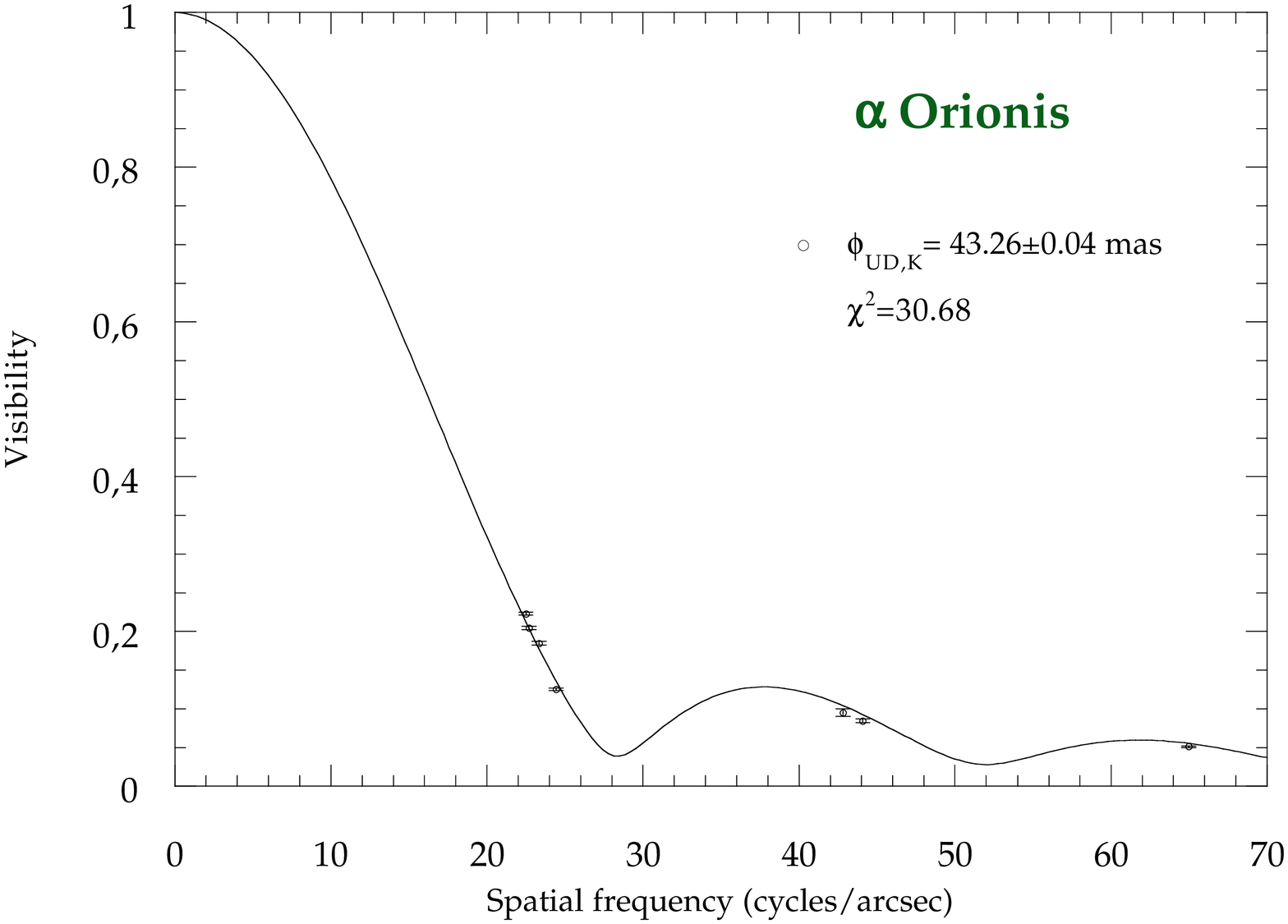}
      \caption{K band visibilities for $\alpha$~Orionis. The $\chi^{2}$ is computed with first lobe visibilities only.}
         \label{fig:aori}
   \end{center}
     \end{figure*}
\begin{figure*}
    \begin{center}
\includegraphics[width=18cm]{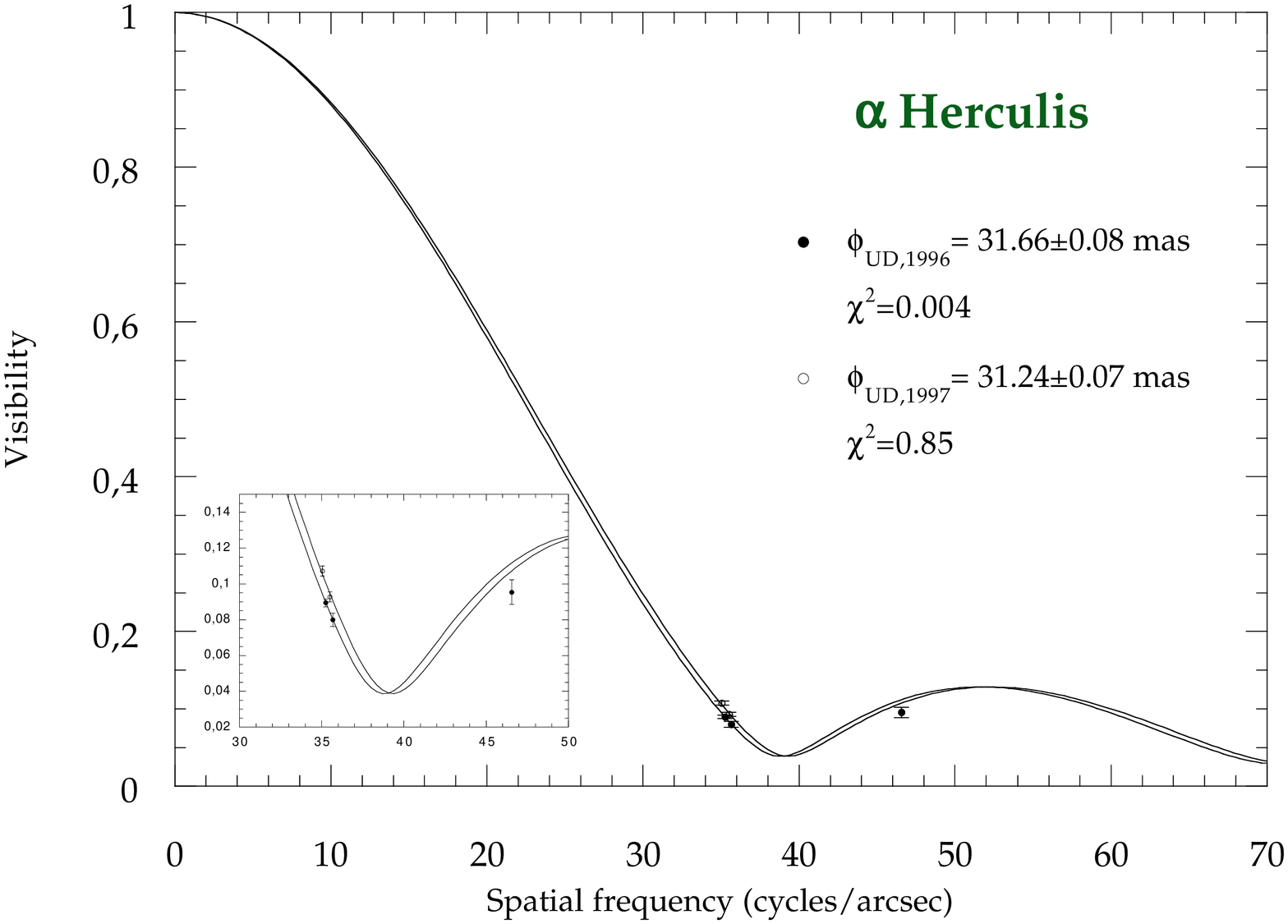}
      \caption{K band visibilities for $\alpha$~Herculis. The $\chi^{2}$ is computed with first lobe visibilities only.}
         \label{fig:aher}
   \end{center}
     \end{figure*}

\subsection{Wide band V$^{2}$ visibility models}
The bandwidth of the K band is about 400~nm inducing a spread of 
spatial frequencies for each individual measurement.  This spread is 
nearly negligible when the average spatial frequency is small, as is 
the case for spatial frequencies in the first lobe of the visibility 
function.  At higher spatial frequencies the spread increases and its 
influence keeps increasing as the visibility model is no longer 
monotonic.  The averaging of visibilities across the bandpass cannot 
be neglected.  Under conditions of turbulence and without a 
fringe tracking system to stabilize differential piston between the 
two pupils, interferograms are distorted and the fringe peak is spread 
leading to a mixing of frequencies.  This raises the issue of whether 
one should average complex visibilities and square the result or 
directly average squared visibilities to compute the wide band 
visibility model. It is not the purpose of this paper to discuss this 
and this point will be tackled in a forthcoming paper. It can be shown 
that for the data discussed here it is legitimate to average 
the squared mochromatic visibilities.\\

Models for which the squared visibility is averaged over the K band 
have been computed.  The monochromatic components are weighted by the 
K band filter transmission and by the spectrum of the source which has 
been modeled by a 3500\,K black-body Planck function.  The chromatic 
splitting ratios of the recombining coupler have also been taken into 
account as wavelengths for which the coupler is 50/50 contribute more 
to the average visibility than those for which the coupler is 
unbalanced.  Splitting ratios have been estimated from narrow band 
measurements.  The maxima of the averaged visibility function are 
smaller than the maxima of a monochromatic visibility model thus 
mimicking a limb darkening effect. Besides the zeroes of the 
monochromatic visibility function are replaced by minima of a few 
percent. It is therefore important to take averaging into account.

\subsection{Model fitting}
The visibility data have been fitted by a wide band uniform disk diameter 
visibility curve by minimizing the function :
\begin{equation}
\chi^{2}=\frac{1}{N-1}\sum_{i=1}^{N}\left(\frac{V_{i}^{2}-M(\O_{UD};S_{i})}{\sigma_{i}}\right)^{2}
\end{equation}
where $\sigma_{i}$ is the estimated error on $V_{i}^{2}$ and:
\begin{equation}
M(\O_{UD};S_{i})=\int_{band} 
{\left|\frac{2J_{1}\left(\pi\phi_{UD}B_{i}k\right)}{\pi\phi_{UD}B_{i}k}\right|}^{2}w(k)dk
\end{equation}
is the wide band uniform disk model with $k$ the wavenumber, 
$S_{i}=B_{i}k_{eff}$ the effective spatial frequency, $B_{i}$ the 
projected baseline and $\O_{UD}$ the uniform disk diameter, 
$w(k)$ being a weighting function.\\

If the model is a perfect representation of the source and if the 
error bars are well estimated then the mean of $\chi^{2}$ is equal to 
1.  The value of the $\chi^{2}$ can therefore be used as a criterion 
to assess the validity of the error bar estimates.\\

The K band visibility data of $\alpha$~Orionis have been gathered and 
fitted by a single visibility function.  The 1996 and 1997 K band data 
for $\alpha$~Herculis are fitted separately.  We present the results 
of the fits for the first lobe data only (spatial frequency $\leq$ 30 
and 40~cycles/arcsec for $\alpha$~Orionis and $\alpha$~Herculis 
respectively) and for all data points.  The results are the following 
for all epochs:
\begin{tabular}{lll}
               \noalign{\smallskip}
            \noalign{\smallskip}
    $\alpha$~Orionis \\
    $\O_{UD,1st}=43.26\pm0.04$\,mas & $\chi^{2}=30.68$  \\
    $\O_{UD,all}=43.33\pm0.04$\,mas & $\chi^{2}=21.45$  \\
               \noalign{\smallskip}
            \noalign{\smallskip}
 
     $\alpha$~Herculis \\
     $\O_{UD,1996,1st}=31.66\pm0.08$\,mas & $\chi^{2}=0.004$  \\
     $\O_{UD,1996,all}=31.64\pm0.08$\,mas & $\chi^{2}=3.37$  \\
     $\O_{UD,1997}=31.24\pm0.07$\,mas & $\chi^{2}=0.85$  \\
                    \noalign{\smallskip}
            \noalign{\smallskip}
\end{tabular}


The data are presented in Figures~\ref{fig:aori} and \ref{fig:aher}, 
with the uniform disk model fits.  The agreement of the first lobe 
data of \her to the model is virtually ``perfect'', setting the upper 
level of accuracy of the FLUOR data to 0.2\%, and confirming that the 
measurement error bars are reasonable.  This conclusion has also been 
verified on other bright stars. Despite the correctness of the 
error bar estimates, the accuracy of the diameter measurements may be 
questionable.  As a matter of fact the diameter estimates depend on 
the estimated effective wavelength of the instrument and are directly 
proportional to it.  The direct measurement of the effective 
wavelength is quite difficult and depends on a large number of 
parameters.  We have assessed the quality of our diameter estimates by 
comparing them to some independent measurements in the paragraph 4.5.3 
of \cite{perrin98}.  The diameters of $\alpha$~Boo and $\alpha$~Tau 
measured with an accuracy of 1 part in 200 with FLUOR were 
statistically compatible with measurements obtained at the same or 
different wavelengths with other interferometers or by the lunar 
occultation technique.  In the case of \ori and \her we are therefore 
confident in our diameter estimates at least at the level of 1 part in 
200. With this understanding, the data in Figures~\ref{fig:aori} 
and \ref{fig:aher} suggest or confirm several conclusions.  \\

First, the consistency of the observations and the small error bars with
the simple model suggests that the surface of \her is well behaved, in some
sense - there is no evidence here for a substantial surprise. \\

Second, there has probably been a slight change of apparent diameter of
0.4\,milli-arcsec between April 1996 and March 1997 as
Figure~\ref{fig:aher} shows. This is evidently a small effect, but is
clearly indicated by the data at the level of precision which is believed
to apply.  For such a small effect however, it is not clear if the
photospheric radius has actually changed, or if an opacity or temperature
variation has changed the brightness profile slightly. \\

Third, in the case of Betelgeuse, we immediately see that the data do 
not fit the simple model as well. There may be a calibration 
problem.  Yet, we wish to suggest a possible alternative explanation 
to this issue. It appears that the surface of the star cannot be 
considered smooth and the brightness distribution must have some 
roughness.  In fact such an effect has been predicted to be in the 
range of 0.1 to 1\% (\cite{ovdl}) depending on the spectral type of 
the star.  This roughness of the visibilities may thus induce large 
values of $\chi^{2}$, and the fit may be very sensitive to which 
visibility values are included.  To illustrate this, the point with 
S~=~24.43~cycles$/$arcsec seems slightly inconsistent with the three 
other first lobe data points.  Without this visibility point, the 
$\chi^{2}$ of the fit with the first lobe points is 7.70 and becomes 
15.00 if second and third lobe data points are added.  The statistics 
are thus significantly changed and are therefore fragile.  \\

The fourth conclusion to be drawn is that the full range of visibility data
for each star is in neither case consistent with a uniform disk model. This
means that limb darkening has to be taken into account, as discussed in the
next section. \\


\section{Physical diameters}
\begin{figure*}
    \begin{center}
\includegraphics[width=18cm]{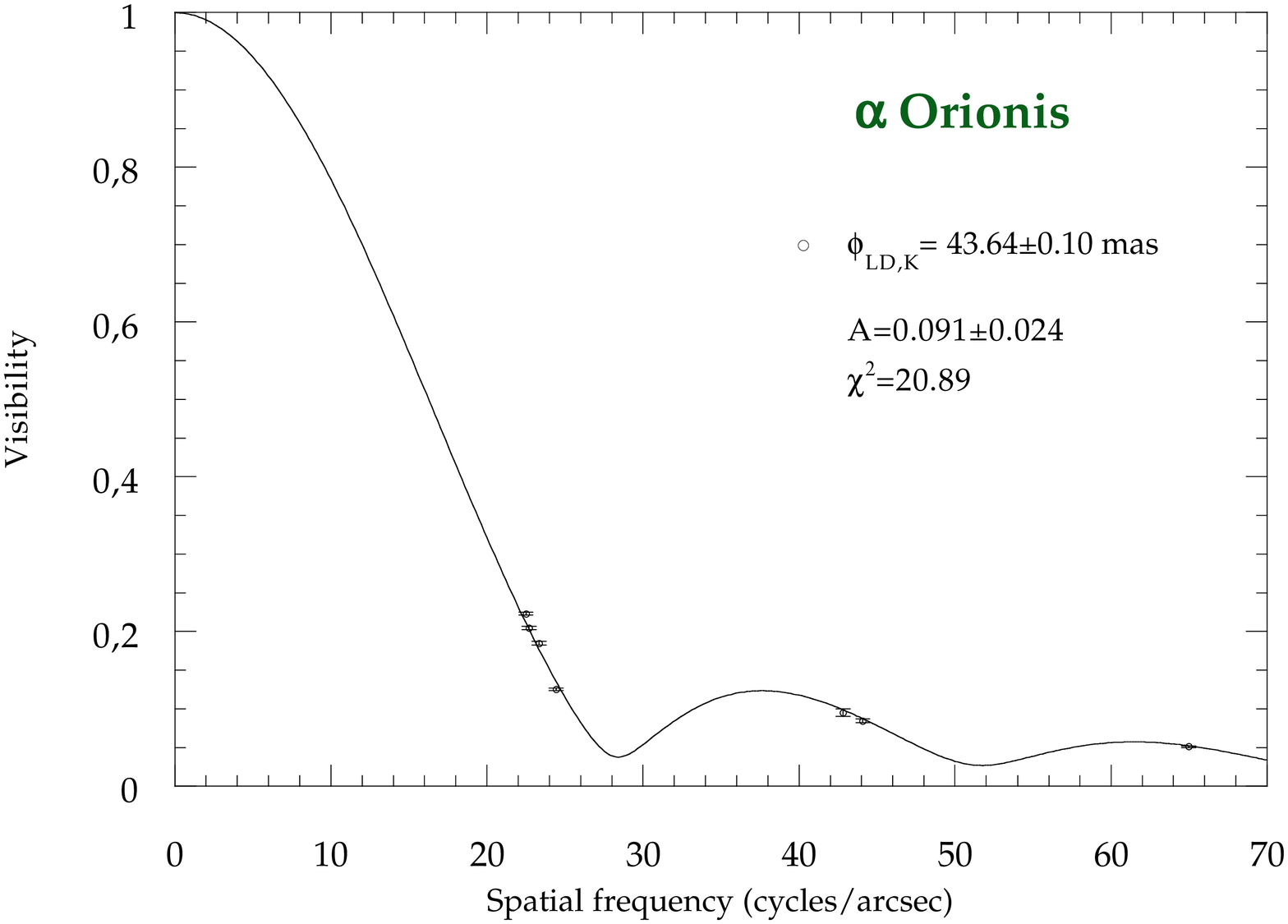}
      \caption{Fit of \ori data by a limb darkened disk model}
         \label{fig:aoriLD}
   \end{center}
     \end{figure*}
\begin{figure*}
    \begin{center}
\includegraphics[width=18cm]{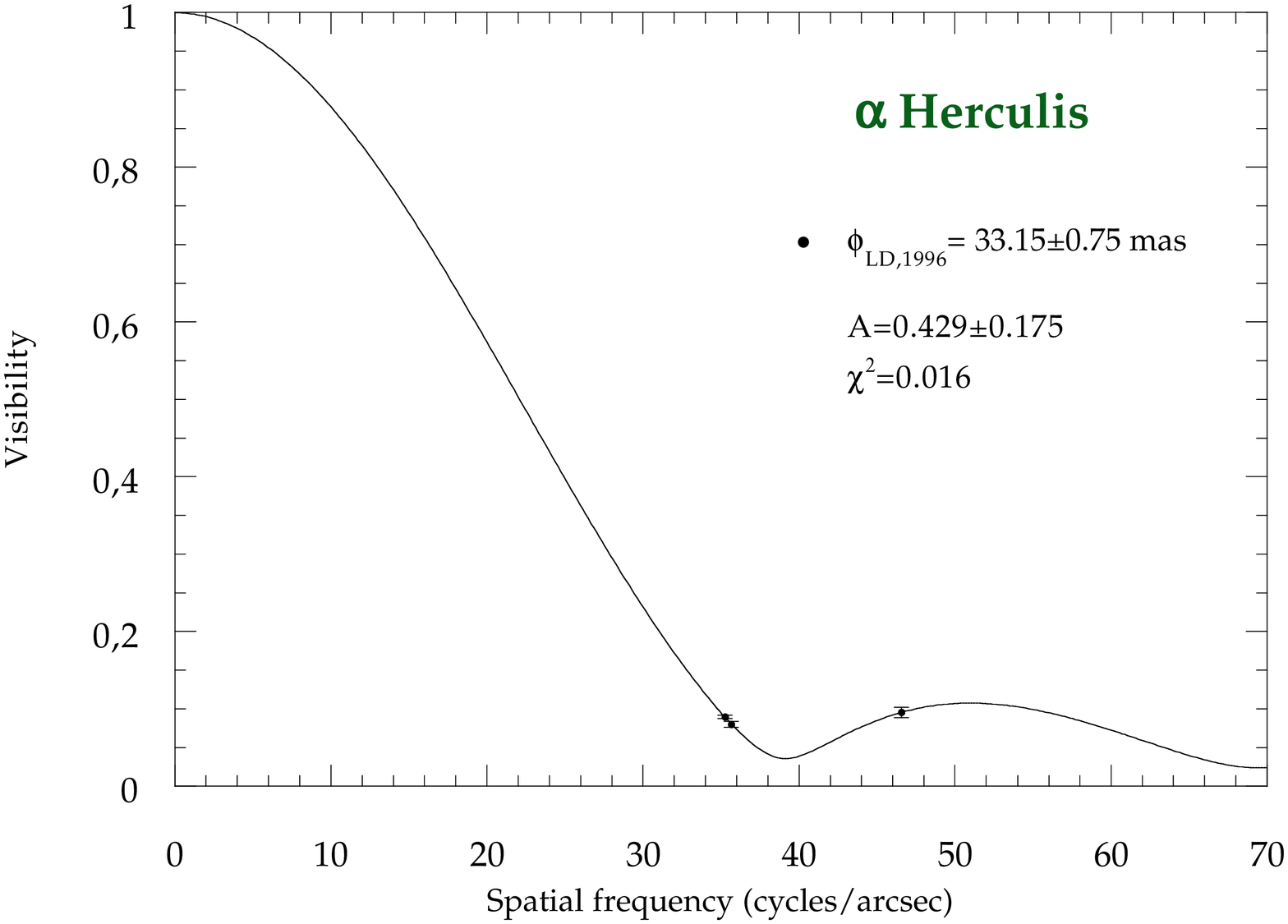}
      \caption{Fit of 1997 \her data by a limb darkened disk model}
         \label{fig:aherLD}
   \end{center}
     \end{figure*}

Discrepancies between the uniform disk and limb darkened disk models 
mostly occur after the first zero of the visibility function. Data in the K 
band have been obtained in this range of spatial frequencies and must 
take limb darkening effects into account. Wide band chromatic effects 
need also to be taken into account to properly estimate the limb 
darkening parameters.

\subsection{Limb darkening measurements on \ori and \her}
\label{sec:LD}
Several limb darkening models have been published. With a limited number 
of data points available for the two stars, only single 
parameter models are used. We have adopted the classical linear limb 
darkening model and the power law model as proposed by 
\cite{hestroffer97}. These can be expressed as:
\begin{eqnarray}
    I(\mu) & = & 1-A(1-\mu)
    \label{eq:linear}  \\
    I(\mu) & = & {\mu}^{\alpha}
    \label{eq:power}
\end{eqnarray}
where $\mu$ is the cosine of the angle between the line of sight and 
the vector joining the current point to the center of the star. The 
intensity is normalized with respect to the center of the stellar 
surface. \\
\subsubsection{\her}
Only the 1996 data are used since no data were taken in the second 
lobe in 1997. The results of the fit are listed in the table below: 
\\ \\
\begin{tabular}{cccc}
    \hline
    Model & Diameter (mas) & Parameter & reduced $\chi^{2}$  \\
    \hline
    Power & $33.59\pm0.89$ & $0.394\pm0.180$ & 0.016  \\
    Linear & $33.14\pm0.76$ & $0.429\pm0.176$ & 0.016  \\
    \hline
\end{tabular}
\\ \\
The fitting residuals are smaller than with a uniform disk model and 
both limb darkening models yield the same diameter. The limb 
darkening parameter is clearly determined in both cases. There is a 
6\% increase between the uniform disk model and the limb darkened disk 
model. \\
Unfortunately no physical value for the $\alpha$ parameter of the 
power law has been predicted. For the linear law, \cite{vanhamme93} 
has predicted a coefficient of 0.321 for a star with an effective 
temperature of 3500\,K and a surface gravity of 
log\,g=0.5. For the same set of physical parameters, \cite{claret2000} 
predicts a coefficient of 0.436. Our measurement is therefore in 
excellent agreement with both predictions, but cannot distinguish between
them.
\subsubsection{\ori}
The same two models were applied to the \ori data. The results are 
reported below:\\ \\
\begin{tabular}{cccc}
    \hline
    Model & Diameter (mas) & Parameter & reduced $\chi^{2}$  \\
    \hline
    Power & $43.76\pm0.12$ & $0.071\pm0.018$ & 20.7  \\
    Linear & $43.65\pm0.10$ & $0.090\pm0.025$ & 20.9  \\
    \hline
\end{tabular}
\\ \\
The results of the fits are displayed on Figure~\ref{fig:aoriLD} and 
\ref{fig:aherLD}. The reduced $\chi^{2}$ are lower than the one 
obtained with a uniform disk model showing that taking into account a 
limb darkening effect improves the fitting of the data.  Both 
parameters are well constrained by the data.  Yet, the coefficient of 
the linear model is much smaller than the ones predicted by the above 
cited authors as opposed to what has been found for \her$\!\!\!$.  The surface 
of Betelgeuse would therefore be far less limb darkened than that of 
\her$\!\!\!$, which may be surprising for stars with similar spectral types.  
The increase in physical diameter due to the limb darkening effect is 
only of 1\%.  Errors on the physical parameters of Betelgeuse can 
hardly be invoked as the linear law coefficient computed from models 
does not seem to vary much in a wide range of surface gravity and 
temperature values.  A calibration error in the data is not a 
plausible explanation.  The highest frequency point in the first lobe 
is obviously introducing the largest error in the fit and is very 
constraining as it is located closer to the null.  Removing this point 
doubles the linear coefficient and does not change the diameter.  In 
any case this is not sufficient to change the conclusion.  Assuming 
that the number of points is sufficient to constrain well the single 
parameter limb darkening profile of Betelgeuse, we report that the 
effect is much smaller in the K band than predicted.

\subsection{Linear radii}
Using the HIPPARCOS parallaxes published by \cite{perryman97} linear 
radii can be derived for both stars (we have used the formulas by 
\cite{browne02} to compute the error for the ratio of gaussian 
distributions):
\begin{eqnarray}
    R_{\mbox {\scriptsize \her}} & = & 460\pm130\,R_{\odot}
    \label{eq:}  \\
    R_{\mbox {\scriptsize \ori}} & = & 645\pm129\,R_{\odot}
    \label{eq:}
\end{eqnarray}
The errors on the parallaxes are unfortunately large and lead to 
inaccurate linear values for the stellar radii. Yet, these values could 
also be used to differentiate supergiants and bright giants from 
regular giants. Giants with the same spectral types would have radii 
of a few tens of solar radii. Supergiants and bright giants therefore 
have radii an order of magnitude larger. 

\section{Effective temperatures and luminosities}
Bolometric fluxes for \ori and \her have been determined by several 
authors and are generally compatible. We have performed estimates of the bolometric fluxes independent 
of these earlier values. Data to achieve this are sparse and from 
different sources. Methods to perform the calculation are also 
different. It is therefore interesting to get independent values.
We have computed bolometric 
fluxes from infrared data listed in the 1999 edition of the 
\cite{gezari93} catalog available at CDS. The data have been 
complemented by UBVR data from the Simbad database.  The data have 
been fitted by a blackbody Planck function to derive a bolometric 
flux.  In the case of \ori$\!\!\!$, data above 5\,$\mu$m have been 
rejected in computing the temperature, as they are dominated by 
circumstellar dust emission - the flux reradiated by the dust is added 
back in to the total luminosity below, by estimating the energy lost 
to absorption.  No clear circumstellar dust emission is detected in 
the case of \her and all available data have been used.  In order to 
derive a realistic error bar on the bolometric flux, we have 
associated an ad-hoc common error bar to each photometric measurement 
by forcing the reduced $\chi^{2}$ to 1.  Two parameters are 
constrained by this fitting procedure: a scaling factor proportional 
to the total stellar flux and a temperature.  Errors computed on these 
two parameters are used to derive an error on the bolometric flux by a 
Monte-Carlo method.  This procedure takes into account the average 
measurement uncertainty and the photometric variability of the object 
as the measurements span different epochs.  \\

Although the two sources are quite close, evidence for interstellar 
reddening has been investigated.  There is no significant influence of 
the interstellar medium below 80\,pc.  We have used the survey by 
\cite{perry82} up to 300\,pc and rescaled them to distance to assess 
the amount of reddening in volumes close to \ori and \her$\!\!\!$.  There is 
no significant reddening for Betelgeuse, the maximum visual extinction 
being on the order 0.04 after eliminating inconsistent values. The 
same procedure was applied to \her and a mean visual  extinction of 
0.128 was adopted. \\
Evolved stars may also have circumstellar dust causing reddening.  
There is no evidence of reddening for \her as no infrared excess is 
detected from the IRAS fluxes and since the $\left[ \mbox{B-V} 
\right]$ color index is compatible with the intrinsic color of an M5 
supergiant given in \cite{johnson66}.  We have assumed that the source 
of extinction is diffuse interstellar dust and we have used the 
extinction law of \cite{mathis90} to compute extinction at any 
wavelength which gives a ratio of 1.044 between the de-reddened and 
the reddened bolometric flux for a temperature of 3300\,K. \\
In the case of Betelgeuse, circumstellar dust signatures are clearly 
visible above 10\,$\mu$m and reddening by the circumstellar dust needs 
to be corrected.  We have applied the procedure used in \cite{dyck92} 
who derive a visual extinction of 0.5. This results in a factor of 
1.203 between the de-reddened and reddened bolometric fluxes.  \\
The bolometric fluxes and associated errors are listed in the table 
below.  We end with values very close to those used by \cite{dyck92} 
and \cite{benson91} which were of 40.9 and 
108.3$\times 10^{-13}\,{\mbox{Wcm}}^{-2}$ for \her and \ori respectively.  
\\ \\
The effective temperature is determined by assuming that the star 
radiates as a black body and has a physical diameter given by its limb 
darkened disk diameter.  The effective temperature is then:

\begin{equation}
    {\mbox{T$_{\mbox{\scriptsize eff}}$}}=7400\,{\left( 
    \frac{{\mbox{F$_{\mbox{\scriptsize bol}}$}}}{10^{-13}\,{\mbox{Wcm}}^{-2}} 
    \right)}^{1/4}{\left( 
    \frac{1\,{\mbox{mas}}}{\O_{\mbox{\scriptsize LD}}}\right)}^{1/2}\,\,{\mbox{K}}
    \label{eq:teff}
\end{equation}

The effective temperature estimates derived from the diameters and 
bolometric fluxes and luminosities are listed in the following table:\\ \\
\noindent 
\begin{tabular}{cccc}
    \hline
    Star & F$_{\mbox{\scriptsize bol}}$  & 
    T$_{\mbox{\scriptsize eff}}$ & Log$\frac{{\mbox L}_{\scriptsize 
    \star}}{{\mbox L}_{\odot}}$\\
         & ($10^{-13}$Wcm$^{-2}$) & K & \\
    \hline
    \ori & $111.67\pm6.49$ & $3641\pm53$ & $4.80\pm0.19$ \\
    \her & $42,62\pm4.15$ & $3285\pm89$ & $4.30\pm0.30$\\
    \hline
\end{tabular}
\\ \\

The luminosity is exactly as expected from \cite{allen2000} for 
the supergiant \ori but is much too low for \her$\!\!\!$.  The discrepancy for 
\her is probably due to a lack of data for this very red supergiant 
class as the luminosity estimate reported in \cite{allen2000} for an 
M5 type seems inconsistent with that for an M2. We have adopted the 
limb darkened disk diameter values derived from the linear model.  The 
effective temperature of both stars is systematically low by 
100-150\,K compared to the effective temperature expected for giants 
with similar spectral types (see e.g. \cite{perrin98}).  Such an 
effect has already been reported (\cite{dyck92}).  The magnitude of 
the effect is here shown to be somewhat smaller than seen in previous 
work with classical beam combiners.  This systematic effect might be 
either due to a systematic error on the diameter which should then be 
underestimated by 7-8\% or on the bolometric flux which should be 
overestimated by 15-17\%.  Such a big difference on the bolometric 
flux does not seem likely.  Errors may arise from the computation of 
the de-reddened flux.  Yet, in the case of \her$\!\!\!$, the correction is 
only 4\% and cannot be invoked to explain this effect.  An 
underestimation of the limb-darkening by 7\% would require a huge 
amount of darkening which is not predicted by the models and is not 
detected in the visibility data.  Besides, limb-darkening coefficients 
decrease with temperature and an increase in limb-darkening should be 
inconsistent with an increase in effective temperature.  We therefore 
think that the temperature of these supergiants (or bright giants) is 
systematically lower than that of giants of same spectral type.\\

The diameters we find are among the smallest in the range of 
diameters measured so far (Figure~\ref{fig:UDvslambda}).  This 
suggests that the K band measurements see deeper in the atmosphere 
than the measurements at longer or shorter wavelengths.  This is 
qualitatively consistent with the fact that the K band is adjacent to 
the 1.6 $\mu$m continuous opacity minimum, though the surprisingly 
large increase in apparent diameter to longer and shorter wavelengths 
remains to be understood and is discussed further below.  Noting that 
since the flux distribution is strongly peaked in the near IR, it 
appears that the apparent size measured at these wavelengths is likely 
to be most representative of the stellar characteristics.  More 
formally, a flux weighted mean diameter would be close to the H and K 
band diameters.  This lends strong support for establishing effective 
temperatures based on these diameter measurements.

\section{Comparison with longer wavelength data}
\label{sec:discussion}
\begin{figure*}
    \begin{center}
\includegraphics[width=18cm]{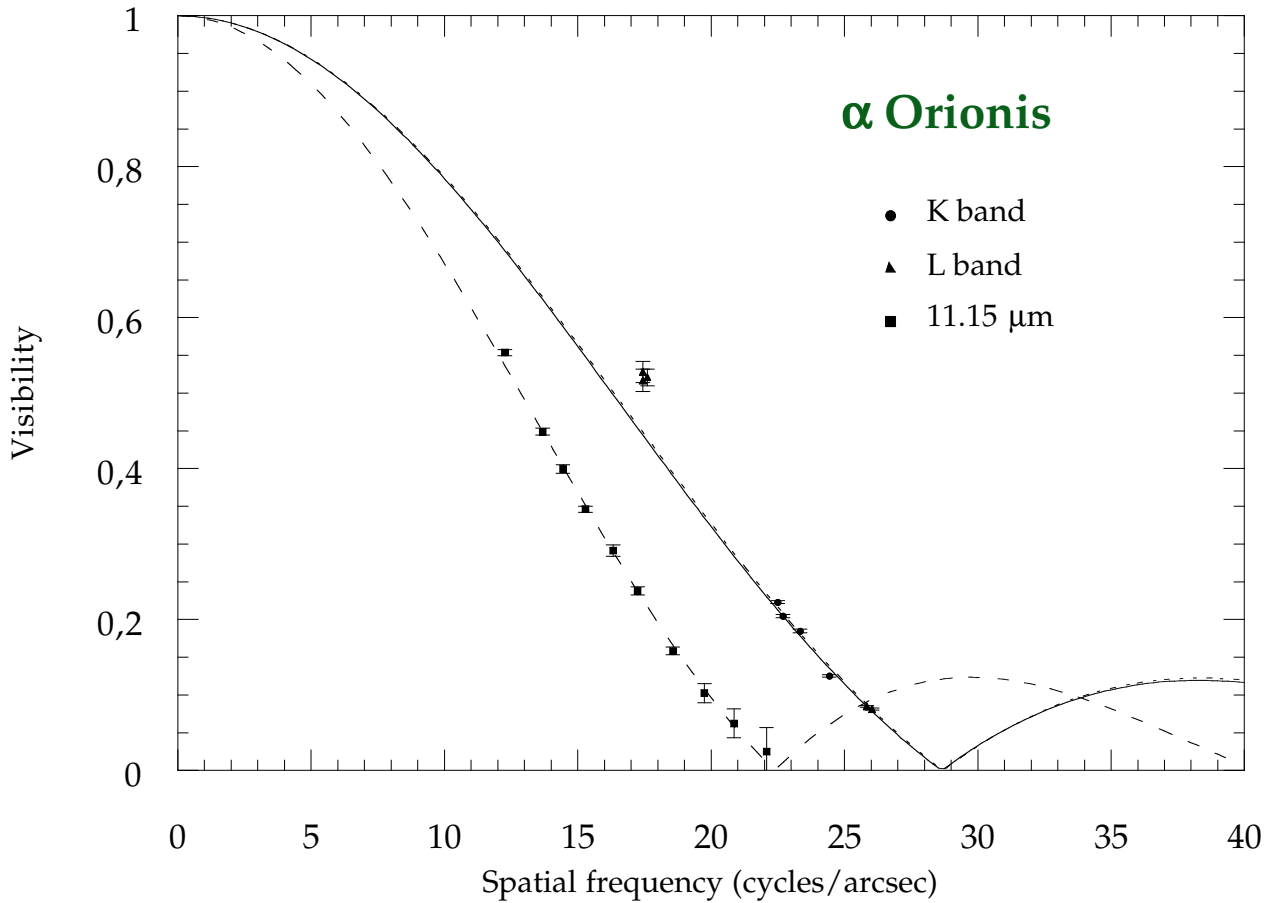}
      \caption{Fit of \ori K, L and 11.15\,$\mu$m data by a 
      photosphere plus layer model. The 11.15\,$\mu$m data have been 
      rescaled to eliminate the contribution of dust (see text). Fits 
      of the K, L and 11.15\,$\mu$m data are respectively the 
      continuous, dotted and dashed lines. Low frequency data in the 
      L band around 18 cycles/arcsec poorly contribute to the fit 
      which is more constrained by the better calibrated data at 26 
      cycles/arcsec.}
         \label{fig:KLN}
   \end{center}
     \end{figure*}
     
The K and L band diameters found by our group are identical to within 
the errors of differential wavelength calibration between TISIS and 
FLUOR. We have used the L band measurement of March 2000 reported 
in \cite{chagnon02}.  The diameter of $42.73\pm0.02$\,mas is 
slightly smaller than the K band diameter but we think this is due to 
the poorer L band calibration and to the much larger uncertainty on 
effective wwavelength.  This situation of very close diameters 
in K and L is dramatically different from what we have found for 
Mira type stars, and can be compared to what could be expected for non 
pulsating giant stars.  Yet, in the case of Betelgeuse, measurements 
at 11.15\,$\mu$m by \cite{weiner2000} show a visibility curve very 
similar to that expected of a photosphere with a diameter 25\% larger 
than the diameter we find at two shorter wavelengths.  Two different 
explanations may be invoked.  The first one would be that the 
limb-darkening might be very strong at shorter wavelengths mimicking a 
much smaller apparent diameter.  This is the explanation of 
\cite{weiner2000}.  However, this is not compatible with the value of 
limb-darkening we find in the K band and with the shape of the 
visibility curve which should be very much altered by such a huge 
effect.\\

Another possibility may be suggested from the study of the atmosphere 
of Mira stars.  We have observed the brightest Mira stars both in the 
L band (\cite{mennesson2002}) and in the molecular and continuum bands 
in K (\cite{perrin2002c}).  We have shown that the visibility data 
could be understood in all bands if a warm molecular layer were added 
around the star.  Because of the huge amount of gas around Mira stars 
(perhaps elevated by dynamical effects associated with pulsation), the 
apparent diameters in the K and L bands are very different and much 
bigger in L due to the lower brightnesss contrast between the 
molecules and the photosphere of the star.  The same feature can be 
tentatively invoked for Betelgeuse to explain the different diameters 
from K and L to 11.15\,$\mu$m.  Narrow band near IR measurements are 
not yet available for Betelgeuse, but we acquired such measurements on 
another supergiant, $\mu$~Cep, which is similar in major respects to 
\ori$\!\!\!$, including evidence for substantial mass loss.  Features similar 
to Mira stars were detected in the visibility measurements of $\mu$~Cep 
(\cite{verhoelst2002}).  This may also apply to Betelgeuse.  As a 
consequence, we have chosen to portray the model as a molecular layer 
in the following, which appears consistent with other evidence.  An 
interpretation in terms of dust would also be possible.  

The dust would require a combination of special properties, including 
high sublimation temperature and/or strongly wavelength dependent 
albedo and emissivity - e.g. \cite{mccabe1982} - in order to survive 
close to the photosphere.  These properties might be possible, but are 
not consistent with currently favored ideas of the nature of 
circumstellar material.  Should dust prove to contribute to the layer 
opacity, this would not change the overall conclusions of this study 
with respect to analysis of the visibility measurements and the 
results for the photospheric dimensions and temperature.\\

We have used the original ISI data at 11.15\,$\mu$m kindly provided by 
J. Weiner.  In their article, the visibilities are modeled by a 
uniform disk visibility function times a constant $A$ smaller than 1 
to take into account the low frequency energy of the dusty 
environment.  The dust is too cold and too far to be seen in the K 
and L bands (\cite{schuller2002}).  We have therefore rescaled the 
original visibilities at 11.15\,$\mu$m by dividing them by the factor 
$A$ which is equivalent to a good approximation in this range of 
spatial frequencies to ignoring the radiation of the dust.  We then 
have searched for a solution to reproduce the K, L and 11.15\,$\mu$m 
data consistently.  We have used a simple shell model of a photosphere 
with a uniform spatial brightness distribution and a spherical layer 
of zero geometrical thickness whose optical thickness is $\tau$.  
$\tau$ is allowed to vary from one band to the other.  The photosphere 
and spherical layer diameters are $\O_{\star}$ and 
$\O_{\mathrm{layer}}$ respectively.  Similarly the respective 
temperatures are $T_{\star}$ and $T_{\mathrm{layer}}$.  The model 
therefore has seven independent parameters.  We have minimized a 
$\chi^{2}$ to find the parameters that best fit the data:

\begin{equation}
    \chi^{2}=\sum_{i=1}^{N}\left[\frac{V_{i}^{2}-M(\O_{\star}, \O_{\mathrm{layer}}, 
    T_{\star}, T_{\mathrm{layer}}, \tau_{\lambda_{i}} ;S_{i})}{\sigma_{i}}\right]^{2}
\end{equation}
where the model $M(\O_{\star}, \O_{\mathrm{layer}}, T_{\star}, 
T_{\mathrm{layer}}, \tau_{\lambda_{i}} ;S_{i})$ is obtained by taking  
the squared Hankel transform of the circularly symetric intensity distribution 
defined as a function of the wavelength $\lambda$ and of the angle 
$\theta$ between the radius vector and the line between the observer 
and the center of the star :
\begin{eqnarray}
I(\lambda,\theta) & = & 
B(\lambda,T_{\star})exp(-\tau(\lambda)/\cos(\theta)) \\ \nonumber 
& & +B(\lambda,T_{\mathrm{layer}})\left[1-exp(-\tau(\lambda)/\cos(\theta))\right]
\end{eqnarray}
for $\sin(\theta) \leq \O_{\star}/\O_{\mathrm{layer}}$
and:
\begin{equation}
I(\lambda,\theta)=B(\lambda,T_{\mathrm{layer}})\left[1-exp(-2\tau(\lambda)/\cos(\theta))\right]
\end{equation}
otherwise. This simple model is also defined in 
\cite{scholz2001} to illustrate the impact of the circumstellar 
environment on the center to limb variation.  Although it is a quite 
simple view of the atmosphere of an evolved object it is very 
convenient to use as it only depends on a small number of parameters 
and allows relatively easy and quick computations.\\

The visibilities measured in the second and third lobes of the K band
visibility function have not been taken into account for the fit in order
not to bias the fit with a particular limb darkening solution.  As we have
seen previously, the model should  be integrated in the whole band if these
data are to be taken into account.  The model here is computed at the
effective wavelength only and mainly aims at reproducing the wavelength
variations of visibilities.  \\

The hypersurface described by the $\chi^{2}$ is complex and has a 
large number of local minima.  The variations of the hypersurface with 
respect to the optical depth parameters are locally convex so that the 
convergence with respect to these three parameters is very quick.  Our 
algorithm explores the 
$(\O_{\star},\O_{\mathrm{layer}},T_{\star},T_{\mathrm{layer}})$ space 
and finds the optimum set of optical depths for each point in this 
space and then chooses the full set of parameters that lead to the 
smaller $\chi^{2}$.  In order to eliminate non physical solutions we 
have first linked $\O_{\star}$ and $T_{\star}$ to keep the bolometric 
flux emitted by the photosphere constant.  These two parameters were 
then set independent for the search of the final solution.  The 
uncertainties on the parameters were computed by varying the optimum 
$\chi^{2}$ by 1.  The values we have found for the best fit parameters 
are listed below: \\

\begin{tabular}{ll}
$\O_{\star}=42.00\pm0.06$\,mas ,  & $T_{\star}=3690\pm50$\,K  \\
$\O_{\mathrm{layer}}=55.78\pm0.04$\,mas , & $T_{\mathrm{layer}}=2055\pm25$\,K \\
$\tau_{K}=0.060\pm0.003$ & \\
$\tau_{L}=0.026\pm0.002$ & \\
$\tau_{11.15\mu m}=2.33\pm0.23$ &\\
\end{tabular}
\\ 

\noindent The best fit visibility curves are represented on 
Figure~\ref{fig:KLN}.  The model reproduces the observed K and 
11.15\,$\mu$m data quite well.  The values of the parameters are 
interesting.  Qualitatively, a star appears larger when seen through 
an absorbing layer with thermal emission.  Consequently, we find a 
smaller diameter for the photosphere of Betelgeuse than with a 
classical limb-darkened disk model.  The effective temperature is 
therefore larger and compatible with the temperature of an M1.5 star.  
 The diameter of the photosphere is mostly constrained by the K 
and L band band data whereas that of the layer is constrained by the 
ISI data.  The apparent diameters in K and L are adjusted with the 
optical depths at these wavelengths.  Their magnitude being small, the 
apparent diameter of the star is very sensitive to them hence their 
excellent statistical accuracy.  This statistical accuracy is 
physically questionable as we think the L band data are not as well 
calibrated as the K band data.  On the contrary, the 11.15\,$\mu$m 
optical depth being so large it is not so well constrained as the 
apparent diameter of the star at this wavelength is not as sensitive 
to it. Taking this model into account, the set of fundamental 
parameters for Betelgeuse is therefore:
\begin{eqnarray}
    R_{\mbox{\scriptsize \ori}} & = & 620\pm 124\,R_{\odot} \\ \nonumber
    T_{\mbox{\scriptsize \ori}} & = & 3690\pm 50\,K
\end{eqnarray}
\noindent The layer is found less than half a stellar radius above the 
photosphere and its temperature is about $2000\,$K.  The physical
significance of the temperature is not clear, since the nature of the
associated opacity is not certain.\\
Of course with a several parameter model, it may not be surprising that
a fit to several data sets could be obtained.  To underline the plausibility of
this model, it is also important to appeal to other possibly related
results.  Numerous studies have given evidence for a high molecular atmosphere
around Betelgeuse.  We note that our derived temperature, $2000\,$K, 
is in agreement with the temperature of $1500\pm500$\,K of a water 
layer described by \cite{tsuji2000}.  Unfortunately no distance for 
the water layer could be determined by Tsuji.  The optical depths in the 
K and L band are very small whereas it is quite important at 
11.15$\,\mu$m. Water vapor modeled for temperatures smaller than the 
photospheric temperature is an absorbant at 11.15\,$\mu$m whereas its 
absorption can be neglected at the temperature of the 
photosphere~(\cite{decin2000}), \cite{weiner2000} only considered this 
last possibility to conclude that water vapor opacity was negligible 
in their study. Silicon monoxide is also a contributor to opacity at 
11.15\,$\mu$m and is negligible in K and L. The contribution of water 
vapor to opacity is also smaller at shorter wavelengths in the K and 
L bands. Thus molecules may be a strong candidate for the shell opacity.

The consistent modeling of interferometric measurements of \ori in the K
and L bands and at 11.15\,$\mu$m is the primary new result presented here.
The success of the shell model in understanding the relation of near-IR and
mid-IR measurements may be qualitatively compatible with the short
wavelength measurements also.   We note first that the visible and mid-IR
apparent diameters are similar. It appears difficult to escape the
conclusion that the atmosphere of \ori is greatly extended, as in the case
of the mira stars.  In fact, this result is not unexpected, as it has been
seen with lunar occulation observations of the M supergiant $\alpha$ Sco
(\cite{schmidtke89}). The upper atmospheric layers (which we approximated as a
thin shell) likely have both scattering and absorptive opacities which vary
with wavelength.  Coupled with the dependence of the Planck function on
wavelength, a rich range of appearances may be observed.  The spotted
appearance seen at short wavelengths may scarcely appear in the infrared,
due to the reduced Plank contrast.  

Since the brightness distribution across \ori is almost certainly not 
properly described with a simple uniform or darkened disk model, the 
limb darkening derived in section~\ref{sec:LD} does not necessarily 
describe the brightness distribution of the photospheric layer, and it 
may not be applicable to the photospheric component alone.  In spite 
of the uncertainty in the limb darkening, it is already clear that the 
shell model leads to an effective temperature which is consistent 
within measurement errors with the Teff of giants of similar spectral 
type.  We are not able at this time to conclude that the low apparent 
Teff of supergiants is an artifact of the outer atmospheric layers, 
but it must be admitted as a possiblity.  A similar multiwavelength 
analysis of additional stars will be needed (for \her$\!\!\!$, the required 
mid-IR data is not available).  Eventually, of course, it will be 
necessary to interpret such measurements in the context of models 
which successfully reproduce the large atmospheric extent.

\section{Conclusions}
 We have been able to measure accurate diameters in the K band for 
 $\alpha$~Herculis and Betelgeuse.  The quality of the calibration of 
 the visibilities sets the current limit of the absolute accuracy with 
 a single-mode fiber interferometer like FLUOR to 0.2\%.  A slight 
 change of diameter of 0.4\,mas for \her has been measured between 
 1996 and 1997, but it is impossible to guess the nature of this 
 diameter change which may be due either to a physical increase or to 
 a change of opacity of the atmosphere.  Visibility data acquired 
 above the first null of the visibility functions allowed us to 
 constrain linear and power law limb-darkening coefficients.  The 
 value found for $\alpha$~Herculis is in agreement with predictions 
 whereas the disk of \ori shows very little darkening. The diameters 
 measured are among the smallest in the observed range of wavelengths, 
 most likely meaning that we see deeper in the atmosphere of these 
 stars in the K band.  
 The data are compatible, although they are sparse, with circular 
 symmetry and there is no clear evidence of departure from circular 
 symmetry due to the presence of spots detected at shorter 
 wavelengths.
 This is in agreement with the result from 
 \cite{young2000} who found no spot signature at 1290\,nm which can be 
 explained by a lower contrast with an increasing wavelength between 
 the Planck function brightness of the photosphere and of the 
 atmosphere.  A consistent analysis of the \ori data in the K and L 
 bands and at 11.15\,$\mu$m has been conducted.  The three sets of 
 data can be explained with a model comprising a $3690\,$K photosphere 
 and a $2055$\,K warm layer located 0.33\,$R_{\star}$ above the 
 photosphere.  This analysis strengthens our interpretation of the 
 direct measurement of the photospheric diameter in the K band and 
 establishes for \ori an effective temperature consistent with the 
 giant stars of the same spectral type.  A consistent scenario to 
 explain the observations of this star from the visible to the 
 mid-infrared can be set-up.  The star is seen through a thick, warm 
 extended atmosphere that scatters light at short wavelengths thus 
 slighty increasing its diameter.  The scatter becomes negligible 
 above 1.3\,$\mu$m.  The upper atmosphere being almost transparent in 
 K and L -- the diameter is minimum at these wavelengths where the 
 classical photosphere can be directly seen.  In the mid-infrared, the 
 thermal emission of the warm atmosphere increases the apparent 
 diameter of the star.  \\

 These measurements show that optical interferometry is now well capable of
 exploring the limb darkening and atmospheric extension of bright, high
 luminosity stars. In the
 case of giants and supergiants, it is now possible to critically
 evaluate the adequacy of existing models to account for the atmospheric
 structure, and to determine for which stellar temperatures and
 luminosities the use of static models breaks down.  With the increasing
 capabilities of optical interferometry, we can expect future measurements
 to record spatial measurements with higher spectral resolution and
 spectral coverage, better evaluating opacity sources and probing the depth
 dependence of the atmospheric conditions, and directly observing stellar
 atmospheric inhomogeneities and variations.

\begin{acknowledgements}
The authors first wish to thank T.~Verhoelst for pointing to the work of 
L.~Decin. The authors are also very grateful to J. Aufdenberg for his careful reading of the paper and for his constructive comments. Lastly, the authors are indebted to the referee whose interactions have improved the quality of this work. 

\end{acknowledgements}

\end{document}